\documentclass[12pt]{article}		

\usepackage[super,sort,comma]{natbib}

\usepackage{fancyhdr}		

\usepackage[section]{placeins}   %

\usepackage{graphicx}

\usepackage{xcolor}

\usepackage{amsfonts}

\usepackage{amsmath}

\usepackage{siunitx}

\usepackage{multirow}

\usepackage{makecell}

\usepackage{hyperref}
\hypersetup{ colorlinks,
    citecolor=blue,
    filecolor=blue,
    linkcolor=blue,
    urlcolor=blue
}

\usepackage[all]{hypcap}    

\makeatletter \renewcommand\@biblabel[1]{$^{#1}$} \makeatother
\newcommand{\captionv}[3]{\begin{center}\parbox{#1}{\caption[#2]{{\sf #3}}}
        \end{center}}

\newcommand{\cen}[1]{\begin{center} #1 \end{center}}

\definecolor{gray}{rgb}{0.6,0.6,0.6}
\definecolor{red}{rgb}{0.85,0,0}
\definecolor{green}{rgb}{0,0.85,0}
\definecolor{blue}{rgb}{0,0,0.85}
\definecolor{beige}{rgb}{0.92,0.87,0.78}

\newcommand{\ie}{{\it i.e.}, }
\newcommand{\eg}{{\it e.g.}, }
\newcommand{\etal}{{\it et al}}

\newcommand{\um}{\si{\micro\meter}{} }
\newcommand{\nm}{\si{\nano\meter}{} }
\newcommand{\mm}{\si{\milli\meter}{} }
\newcommand{\cm}{\si{\centi\meter}{} }
\newcommand{\ums}{\si{\micro\meter}{}}

\newcommand{\cms}{\si{\centi\meter}{}}

\newcommand{\peri}{``peri''{} }
\newcommand{\oendo}{``1-endo''{} }
\newcommand{\fendo}{``4-endo''{} }
\newcommand{\peris}{``peri''{}}
\newcommand{\oendos}{``1-endo''{}}
\newcommand{\fendos}{``4-endo''{}}

\newcommand{\del}{$\%\Delta_\text{DEF}${} }
\newcommand{\delp}{$\%\Delta_\text{DEF}^+${} }
\newcommand{\delm}{$\%\Delta_\text{DEF}^-${} }
\newcommand{\dels}{$\%\Delta_\text{DEF}${}}
\newcommand{\delps}{$\%\Delta_\text{DEF}^+${}}
\newcommand{\delms}{$\%\Delta_\text{DEF}^-${}}

\newcommand{\cellDEFs}{n,cDEF{}}

\begin{document}

\cen{\sf {\Large {\bfseries Multiscale Monte Carlo simulations of gold nanoparticle dose-enhanced radiotherapy II. Cellular dose enhancement within macroscopic tumor models} \\
\vspace*{10mm}
Martin P. Martinov, Elizabeth M. Fletcher, and Rowan M. Thomson$^\text{a)}$} \\ 
Carleton Laboratory for Radiotherapy Physics, Department of Physics, Carleton University, Ottawa, Ontario, K1S 5B6, Canada \\ \vspace{5mm} 
submitted on \today \\
}

\setcounter{page}{1}
\pagestyle{plain}
email: a) rthomson@physics.carleton.ca \\

\begin{abstract}

\noindent\textbf{Background:} Gold NanoParticle (GNP) dose-enhanced radiation Therapy (GNPT) requires consideration of physics across macro- to microscopic length scales, however, this presents computational challenges that have limited previous investigations.

\noindent\textbf{Purpose:} To develop and apply multiscale Monte Carlo (MC) simulations to assess variations in nucleus and cytoplasm dose enhancement factors (\cellDEFs{}s) over tumor-scale volumes.

\noindent\textbf{Methods:} The intrinsic variation of  \cellDEFs{}s (due to fluctuations in local gold concentration and cell/nucleus size variation) are estimated via MC modeling of varied cellular GNP uptake and cell/nucleus sizes.   Then, the Heterogeneous MultiScale (HetMS) model is implemented in MC simulations by combining detailed models of populations of cells containing GNPs within simplified macroscopic tissue models to evaluate  \cellDEFs{}s.  Simulations of tumors with spatially-uniform gold concentrations (5, 10, or 20~mg$_\text{Au}$/g$_\text{tissue}$) and spatially-varying gold concentrations eluted from a point are performed to determine \cellDEFs{}s as a function of distance from the source for 10~keV to 370~keV photons. All simulations are performed for three different intracellular GNP configurations: GNPs distributed on the surface of the nucleus (perinuclear) and GNPs packed into one or four endosome(s).

\noindent\textbf{Results:}  Intrinsic variations in \cellDEFs{}s can be substantial, \eg if GNP uptake and cell/nucleus radii are varied by 20\%, variations of up to 52\% in nDEF and 25\% in cDEF are observed compared to the nominal values for uniform cell/nucleus size and GNP concentration.
 In HetMS models of macroscopic tumors, sub-unity \cellDEFs{}s (\ie dose {\it decreases})  can occur for low energies and high gold concentrations due to attenuation of primary photons through the gold-filled volumes, \eg \cellDEFs{}$<1$ is observed 3~mm from a 20 keV source for the four endosome configuration.  In HetMS simulations of tumors with spatially-uniform gold concentrations, \cellDEFs{}s decrease with depth into the tumor as photons are attenuated, with relative differences between GNP models remaining approximately constant with depth in the tumor. Similar initial \cellDEFs{} decreases with radius are seen in the tumors with spatially-varying gold concentrations, but the \cellDEFs{}s for all of the GNP configurations converge to a single value for each energy as gold concentration reaches zero.

\noindent\textbf{Conclusions:} The HetMS framework has been implemented for multiscale MC simulations of  GNPT  to compute \cellDEFs{}s over tumor-scale volumes, with results demonstrating that cellular doses are highly sensitive to cell/nucleus size,  GNP intracellular distribution, gold concentration, and cell position in tumor. This work demonstrates the importance of proper choice of computational model when simulating GNPT scenarios and the need to account for intrinsic variations in \cellDEFs{}s due to variations in cell/nucleus size and gold concentration. 

\end{abstract}

\setlength{\baselineskip}{0.7cm}
\pagestyle{fancy}

\section{Introduction} \label{sec_Cell:Introduction}

Gold NanoParticle (GNP) dose-enhanced radiation Therapy (GNPT) is a prospective treatment modality with the potential to enhance dose localization in radiation medicine\cite{He17,Br20}.  With GNPs targeted to tumor tissues, GNPT aims to increase target dose, characterized by a Dose Enhancement Factor (DEF), while also reducing doses of nearby healthy tissues containing few-to-no GNPs.  There are many macro- and microscopic effects in GNPT that  impact GNP effectiveness, \eg GNP uptake on a cellular level\cite{Sa19} or attenuation due to gold\cite{MT17}, which can not be characterized by a single DEF.  Thus, DEF calculations require a multi-factor model that can incorporate effects at microscopic (cell, GNP) scales and the overall effects of introducing a large quantity of gold on a macroscopic (tumor) level \cite{MT17,ZS16}.  

In Part I of this study, Monte Carlo (MC) simulations were used to quantify DEFs on a cellular level, in particular nucleus and cytoplasm DEFs which we collectively denote as \cellDEFs{}s. We established the sensitivity of \cellDEFs{}s across a large parameter space including cell/nucleus sizes, GNP intracellular configuration, gold concentration, and incident photon energy.  This paper, Part~II, builds on that single cell investigation to consider expected variations in  \cellDEFs{}s within macroscopic volumes of interest.  

In realistic scenarios, cell and nucleus sizes are not constant but vary within a population of cells, and, similarly, GNP uptake fluctuates in time and also amongst cells in a cell population \cite{OT17a,Sc16,Ch10d}. As \cellDEFs{}s are sensitive to cell size and GNP concentration (demonstrated in Part~I), these variations/fluctuations will affect \cellDEFs{}s which should not be expected to be constant across a cell population or in time.  Thus, even if mean cell/nucleus sizes and gold concentration in a cell population are known, it is expected that there will be intrinsic variations in \cellDEFs{}s for which we develop estimates herein. 

We then proceed to consider \cellDEFs{}s on macroscopic, tumor scales. To do this, we implement the Heterogeneous MultiScale (HetMS) model \cite{MT17} in conjunction with the models of cells containing GNPs developed in Part~I.  The idea of the HetMS model is to combine {\it heterogenous} models, namely models with varying levels of complexity, on different length scales in a single simulation.  The HetMS approach enables us to model tumor-scale volumes that have either spatially-uniform or spatially-varying macroscopic gold concentrations, and compute \cellDEFs{}s within detailed cell models containing explicitly modeled GNPs, \ie the models developed in Part~I.

To our knowledge, this is the first time that  {\it cellular} (microscopic) dose enhancement factors (\cellDEFs{}s) and their expected variation have been quantified in tumor (macroscopic) volumes of interest through novel multiscale MC simulations.

\section{Methods} \label{sec_Cell:Methods}

\subsection{Monte Carlo parameters}\label{ssec:Methods-general}

Monte Carlo simulations are carried out using the egs\_chamber application of EGSnrc; transport parameters, cross sections, and simulation parameters are described in section II.A of Part~I. This section summarizes some MC simulation methodology used in the present work, with following subsections providing more details. 

Each cell is modeled by two concentric spheres, with the inner sphere representing the nucleus and the region between the nucleus and cell boundaries representing the cytoplasm.  The reference cell, used for most simulations unless otherwise specified, is 7.35~\si{\micro\meter} in radius with a 5~\si{\micro\meter} radius nucleus. The only media used in this work are ICRU four-component tissue (10.1\% hydrogen, 11.1\% carbon, 2.6\% nitrogen, and 76.2\% oxygen by mass\cite{ICRU44}) for all non-gold biological media, pure gold used for GNPs, and a homogeneous mixture of the two materials used for the bulk scatter media.  A hexagonal close-packed lattice\cite{MT20} is used for modeling GNPs clustered within cells, to model a 13 cell cluster (\ie a target cell and its 12 nearest neighbors, see Section~\ref{ssec:Methods-neighbor}), and to place cells within the microscopic scoring regions (see Section~\ref{ssec:Methods-Cyl}). Cells' minimum neighbor distance is 2.06~\si{\micro\metre},  resulting in a cell number density of $3\times10^{8}$ per cubic centimeter for the reference cell size \cite{Th13,OT17}.  

Three different configurations are used to model the GNPs inside a cell containing gold: GNPs clustered closely about the nucleus (\peris) and the GNPs clustered tightly into one (\oendos) or four (\fendos) endosome compartment(s). For a given gold concentration, cells (of the same size) with each of the three configurations of gold all contain an equal mass of gold, but the distribution of the gold differs between configurations. A detailed description and diagram of these configurations is given in Part~I (Section~II.A and Figure~1). 

As described in Sections~\ref{ssec:Methods-Cyl} and~\ref{ssec:Methods-Sphere}, the Heterogeneous MultiScale (HetMS) model\cite{MT17} is used in conjunction with the cell models defined in Part~I to simulate GNPT treatments.  HetMS combines an efficient model for the bulk of the phantom volume with more detailed microscopic models embedded within the phantom in regions of interest to extract cell doses throughout the phantom.  The cell models are used to populate microscopic scoring regions which are then placed into either a cylinder containing a uniform macroscopic gold concentration (section \ref{ssec:Methods-Cyl}) or a sphere containing spatially-varying gold concentrations that decrease as a function of radius (section \ref{ssec:Methods-Sphere}); these two macroscopic models have been used with HetMS previously\cite{MT17}, but were limited by the microscopic model only consisting of GNPs embedded evenly in tissue (as opposed to the more detailed biological cell model used in this work). 

Cell cluster simulations (section \ref{ssec:Methods-neighbor}) are run on the same Intel Xeon 5160 core used in Part~I, and timing is of the same order (less than one to a few hours per configuration). HetMS simulations were performed on the Digital Research Alliance of Canada's Graham cluster:  simulations of uniform macroscopic gold concentration (section \ref{ssec:Methods-Cyl}) took on the order of 10,000 CPU hours per simulation, and  simulations with spatially-varying gold concentrations (section \ref{ssec:Methods-Sphere}) were approximately an order of magnitude faster. Since the HetMS simulations were run on multi-core clusters with CPUs of varying speeds, these times are not representative measures.

\subsection{Intrinsic variation in n,cDEF due to cell size and gold concentration} \label{ssec:Methods-neighbor}

This section describes how we develop quantitative estimates of the intrinsic variations in \cellDEFs{}s due to variations in cell and nucleus sizes in a population of cells, as well as GNP uptake fluctuations amongst cells in a cell population. The first set of simulations quantifies the effect of variation in cellular GNP uptake on \cellDEFs{}s. Simulations are performed for the three gold configurations (\peris, \oendos, and \fendos), nominal gold concentrations of 5, 10, and 20~mg/g, and photon energies from 10 to 370~keV.  Each simulation considers $10^{11}$ histories with no variance reduction techniques. 

\begin{figure}[htbp]
	\centering
	\includegraphics[width=0.35\textwidth]{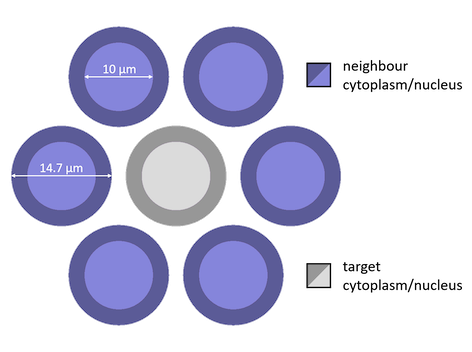}
	\captionv{0.95\linewidth}{13 cell cluster diagram}{Cross-sectional diagram of a central scoring cell (grey) surrounded by all of its neighboring cells (blue) in a close-packed hexagonal lattice configuration. 
	\label{fig:neighborCellDiagram}}
\end{figure}

A cluster of 13 cells comprised of a central target cell and its 12 nearest neighbors (Figure~\ref{fig:neighborCellDiagram}) are modeled at the centre of a 100~\um radius tissue sphere. To model variation in GNP uptake, gold concentration (or, equivalently, GNP number) in the target and neighbor cells are increased/decreased by 20\%, resulting in nine combinations of target and neighbor concentrations, denoted ($c_\text{target}$, $c_\text{neighbor}$), for each nominal GNP concentration.  For example, for the nominal 5~mg/g concentration, the nine combinations of ($c_\text{target}$, $c_\text{neighbor}$) investigated are (4, 4), (4, 5), (4, 6), (5, 4), (5, 5), (5, 6), (6, 4), (6, 5), and (6, 6)~mg/g.  Scenarios for which the gold concentration or cell size would not allow for an endosome that fits within the cytoplasm are omitted from the statistical populations.  From each set of nDEF and of cDEF values corresponding to each given nominal concentration, we extract quartile data, denoted $Q_i^\text{DEF(conc)}$ for the $i$th quartile. We then quantify the upward ($\%\Delta_\text{conc}^+$) and downward ($\%\Delta_\text{conc}^-$) effects of variations due to local gold concentration as:
\begin{equation}
	\%\Delta_\text{conc}^+ = \frac{Q_3^\text{DEF(conc)} - Q_2^\text{DEF(conc)}}{\text{DEF}_\text{ref}} \times 100\% \label{eq:DelDEF1}
\end{equation} \vspace{-0.8\normalbaselineskip}
\begin{equation}
	\%\Delta_\text{conc}^- = \frac{Q_2^\text{DEF(conc)} - Q_1^\text{DEF(conc)}}{\text{DEF}_\text{ref}} \times 100\% \label{eq:DelDEF2}
\end{equation}
where  $\text{DEF}_\text{ref}$ is the nDEF or cDEF of the reference cell with a local gold concentration matching the overall gold concentration.  Values are normalized as a percentage of the reference cell nDEF or cDEF so that they may later be used to determine the range of intrinsic variations.  The inner quartiles are chosen as they significantly reduce the overall variation while still providing a close estimate of the lower bound, discussed in Section~III.D of Part~I.

Similarly, variations in \cellDEFs{}s due to cell size are assessed using statistical data from the DEFs calculated for various permutations of cell and nucleus radius in Part~1 (Section~III.D) are used to determine $(\%\Delta_\text{size}^+,\%\Delta_\text{size}^-)$ in an analogous way to Eqs.~1 and~2.   The quartile data, $Q_i^\text{DEF(size)}$, are taken from the dose enhancement values for sets of cells with sizes ($r_\text{cell}$, $r_\text{nuc}$) = (5, 2), (5, 3), (5, 4), (10, 7), (10, 8), and (10, 9)~\ums, shown in Figure~6 of Part~I.  We consider datasets for each of the three gold configurations (\peris, \oendos, and \fendos), gold concentrations of 5, 10, and 20~mg/g, and source energies between 10 and 370~keV, computing $(\%\Delta_\text{size}^+,\%\Delta_\text{size}^-)$ for each dataset.

Using the estimated variations due to concentration ($\%\Delta_\text{conc}^+,\%\Delta_\text{conc}^-$) and cell size ($\%\Delta_\text{size}^+,\%\Delta_\text{size}^-$), we then calculate the combined intrinsic variations  (\delp,\delm) by adding in quadrature:
\begin{equation}
	\%\Delta_\text{DEF}^+ = \sqrt{(\%\Delta_\text{conc}^+)^2+ (\%\Delta_\text{size}^+)^2}
	\end{equation} \vspace{-0.8\normalbaselineskip}
\begin{equation}
	\%\Delta_\text{DEF}^- = \sqrt{(\%\Delta_\text{conc}^-)^2+ (\%\Delta_\text{size}^-)^2}.
\end{equation}
Thus, there is one \dels = (\delp,\delm) pair for nucleus and for cytoplasm for each of the three gold configurations (\peris, \oendos, and \fendos), gold concentrations of 5, 10, and 20~mg/g, and source energies between 10 and 370~keV, which are compiled to provide ranges of intrinsic \cellDEFs{} over a cell population based on a nominal cell DEF calculation.

\subsection{DEF in a tumor with uniform macroscopic gold concentration} \label{ssec:Methods-Cyl}

A GNP-containing tumor is modeled using the HetMS framework as a 3~\si{\centi\meter} long, 1~\si{\centi\meter} radius cylinder composed of a gold-tissue mixture (concentrations of 5, 10, or 20~mg/g) irradiated by a parallel photon beam (monoenergies of 10-370~keV, as well as $^{125}$I, $^{131}$Cs, and $^{103}$Pd spectra\cite{NCRP58,NNDC}) incident over one of its flat faces.  Microscopic scoring regions defined as 200~\si{\micro\meter} diameter spheres (including the 14.7~\um buffer) are placed at 119 evenly spaced positions along the central cylindrical axis, starting half a spacing from the cylinder face. 

\begin{figure}[htbp]
	\centering
	\includegraphics[width=0.5\textwidth]{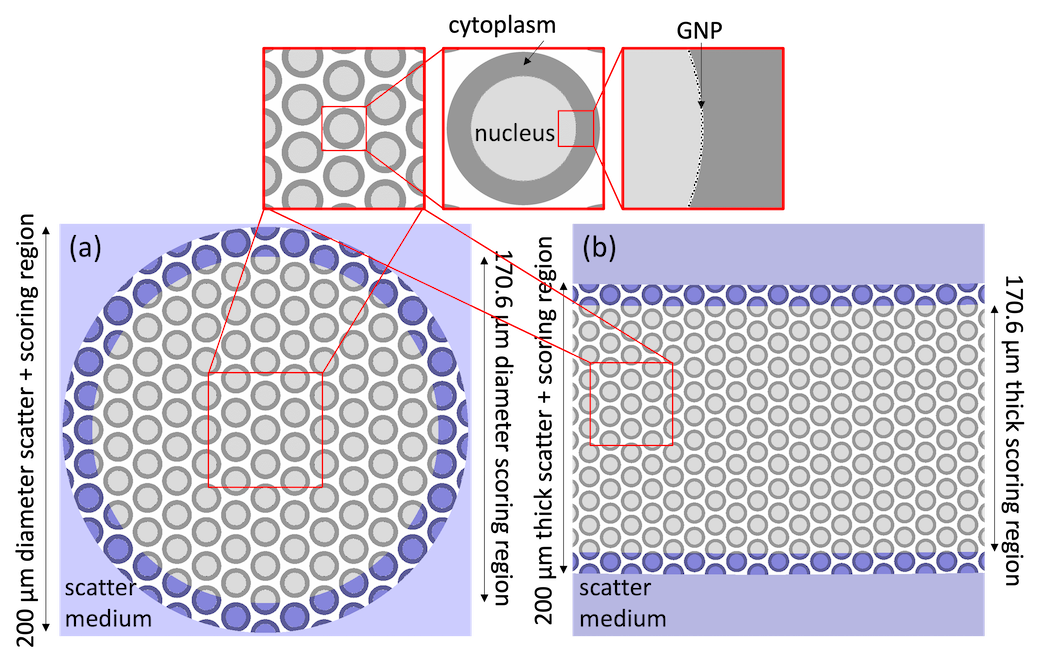}
	\captionv{0.95\linewidth}{Cell microscopic scoring region image}{Microscopic scoring regions for tumor-scale simulations: (a) 200~\um diameter sphere (sec.~\ref{ssec:Methods-Cyl}) and (b) a portion of the 200~\um thick spherical shell (sec.~\ref{ssec:Methods-Sphere}). The diagrams shown depict cells with GNPs in the perinuclear configuration: grey nucleus and cytoplasm are the scoring volumes, blue regions are the buffer (scatter). Surrounding the scoring regions are homogeneous scatter media with matching average elemental compositions.
	\label{fig:microCav}}
\end{figure}

Cells are embedded within each microscopic scoring region by placing them in a hexagonal lattice with 2.06~\si{\micro\metre} spacing between outer radii of nearest neighbours \cite{Th13}.  The total mass of gold in each cell is chosen such that the average mass of gold over mass of tissue in the scoring region matches the gold concentration of the bulk tissue (5, 10, or 20~mg/g), as done in Part~I.  A ``buffer'' region spanning a cell diameter (14.7~\si{\micro\meter}), where cells are still modeled but dose is not scored, is placed along every boundary that interfaces with the bulk homogeneous media for proper local scatter conditions.  Figure~\ref{fig:microCav}a shows a cross-section of a spherical microscopic scoring region and its buffer region embedded in a gold-tissue mixture.

A set of simulations is performed with cell models for each gold configuration (\peris, \oendos, and \fendos).  For comparison, a second set of simulations with GNPs  \textit{dispersed} throughout the scoring cavity are carried out (analogous to previous cubic lattice GNP simulations with no cell models\cite{MT17}): 25~\nm radius GNPs are embedded in a hexagonal lattice pattern to achieve the desired gold concentration and dose is scored to all tissue within the scoring cavity, where the dose enhancement is referred to as dDEF. A set of simulations with only tissue are performed for \cellDEFs{} normalization using two microscopic models: cells containing no GNPs or a cavity filled with only tissue.  Russian roulette with a factor of 64 (1 in 64 chance of survival and survivors have 64 times higher weight) is applied to all electrons that are at a distance from any scoring volume larger than their range under the continuous slowing down approximation.  Simulations are run for $10^{10}$ histories.

\subsection{DEF in a tumor with spatially-varying gold concentration} \label{ssec:Methods-Sphere}

Additional HetMS simulations are performed with a 2.5~\si{\centi\meter} radius sphere containing a GNP eluter and point source in the centre.  This model is based on the work of Sinha \etal\cite{Si15} where the sphere is divided into 51 subregions (a 0.025~\mm thick region, 49 regions with a thickness of 0.05~\mm and another 0.025~\mm thick region), each with a different gold concentration based on the time the eluted GNPs take to diffuse to different radii.  A plot of gold concentration as a function of radius at different times is shown in Figure~3 of the introductory HetMS work\cite{MT17} and is described in more detail in Section~II.C therein. The same time points (1, 5, 33, 59, and 200 days) as in the previous work were simulated here as well, with a focus on the 33 day time point as this scenario has enough time for gold to diffuse multiple scoring shells from the eluter, with enough of a gradient present to see an impact on the observed DEFs. Microscopic scoring regions are defined similarly to those in \ref{ssec:Methods-Cyl}, but rather than spheres they are defined as 200~\um thick shells placed at the centre of each 0.05~\mm thick region (Fig.~\ref{fig:microCav}b). Again, simulations are done with cells for each of three configurations (\peris, \oendos and \fendos) and with \textit{dispersed} GNPs in a hexagonal lattice. A simulation with only tissue (for DEF normalization) is performed with two different microscopic models: cells containing no GNPs or a cavity filled with only tissue.  Typical brachytherapy spectra\cite{NCRP58,NNDC} of $^{125}$I, $^{131}$Cs, and $^{103}$Pd and monoenergic photons with energies between 10 and 370~keV are used in an isotropic point source in the centre of the phantom.  Simulations are run for $10^{10}$ histories and no variance reduction techniques are used.

\section{Results} \label{sec_GNPT:Results}

\subsection{Intrinsic variation in n,cDEF due to cell size and gold concentration} \label{ssec:Results-Range2}

In simulations of the 13 cell cluster, the differences in \cellDEFs{}s when varying the concentration of the 12 neighboring cells by 20\% is found to be very small (0-2\% differences in \cellDEFs{}s) compared to those found when varying target cell concentration (2-21\% differences in \cellDEFs{}s).  The lack of sensitivity to neighbor concentration is the motivating factor in the choice of the one cell diameter buffer region size for the HetMS models, which is chosen to ensure all scoring volumes are within fully modeled cells, though they may be neighboring a partially modeled cell.  Table~\ref{tab:SenseCalc} lists variations when fluctuating cell/nucleus radius and target/neighbor gold concentration for a \peri configuration cell in tissue with a concentration of 20~mg/g and shows a sample calculation of (\delp,\delms).  
\begin{table}[htbp]
	\vspace{0.5\normalbaselineskip}
	\centering
	\captionv{0.95\linewidth}{Example \del variation in GNPT sample calculations}{DEF variation for nucleus, \dels, for a \peri configuration reference cell in 20~mg/g concentration, with contributions due to cell/nucleus size $(\%\Delta_\text{size}^+,\%\Delta_\text{size}^-)$ and target/neighbor concentration fluctuation $(\%\Delta_\text{conc}^+,\%\Delta_\text{conc}^-)$ shown, as well as their quadrature sums (\delp,\delm). 
	\label{tab:SenseCalc}}
	\small
	\begin{tabular}{lccccc}
		\hline \hline \vspace{-0.8\normalbaselineskip} \\
		 \multirow{2}{*}{variation} & \multicolumn{5}{c}{ energy (keV) } \\\cline{2-6}
		 & 20 & 30 & 50 & 70 & 90 \\ \hline \vspace{-0.8\normalbaselineskip} \\
		 $\%\Delta_\text{size}^+$ & 29 & 40 & 11 & 7 & 51 \\
		 $ \%\Delta_\text{conc}^+$& 14 & 14 & 2 & 6 & 13 \\ 
		  \delp & 32 & 42 & 11 & 9 & 52 \\ \hline \vspace{-0.6\normalbaselineskip} \\
		 $\%\Delta_\text{size}^-$ & 22 & 25 & 8 & 7 & 21 \\
		$ \%\Delta_\text{conc}^-$ & 18 & 16 & 7 & 2 & 11 \\  
		 \delm & 29 & 29 & 11 & 7 & 24 \\ \hline \hline 
	\end{tabular}
\end{table}

The set of \delp and \delm values for parameters investigated in this work are compiled in Table~\ref{tab:FluxUp}.  Both \delp and \delm are largest for the \peri configuration, with \del values for nucleus typically larger than those for cytoplasm.  For the \oendo and \fendo configurations, \del values are at or below 10\% for all but a few cases.  The largest \del values occur at energies for which \cellDEFs{} are highest, \ie 20, 30, and 90~keV (shown in Figure~4 in Part~I), and can drop to a few percent at energies for which \cellDEFs{}s are close to unity.  In cases with a \delp of 10\% or higher, the \delp is always larger than or equal to the equivalent \delms, often larger by a factor of 2 or higher; this follows from the ranges found for \cellDEFs{}s in Figure~6 in Part~I. Overall variation in \cellDEFs{}s ranges from a \del of a few percent in a number of scenarios to a \del of (52\%,24\%) for nDEFs in the \peri configuration for a 90~keV beam.

\newpage
\begin{table}[htbp]
	\centering
	\vspace{1cm}
	\rotatebox{90}{\parbox{\linewidth}{
	\captionv{0.95\linewidth}{\del in GNPT}{\del (shown as \delp{},\delm{}) for all configurations, concentrations and (mono) energies investigated.  
	\label{tab:FluxUp}}}}
	\raisebox{.35\textheight}[0pt][0pt]{\rotatebox[origin=c]{90}{
	\centering
	\begin{tabular}{ccrccccccccccc}
	    \hline \hline \vspace{-0.8\normalbaselineskip} \\
	    \multirow{2}{*}{ config. } & \multirow{2}{*}{ \makecell{conc.\\(mg/g)} } & \multirow{2}{*}{ target } & \multicolumn{11}{c}{ energy (keV) } \\
		& & & 10 & 20 & 30 & 40 & 50 & 60 & 70 & 80 & 90 & 100 & 370 \\ \hline \vspace{-0.7\normalbaselineskip} \\
		\multirow{6}{*}{ peri } & \multirow{2}{*}{ 5 } & nuc. & 21,7 & 27,18 & 25,14 & 20,5 & 7,10 & 10,3 & 9,4 & 3,3 & 38,7 & 15,6 & 2,2 \\
		& & cyt. & 3,4 & 13,6 & 4,4 & 2,8 & 4,5 & 2,2 & 4,4 & 5,3 & 8,5 & 6,4 & 1,2 \\ \vspace{-0.5\normalbaselineskip} \\
		& \multirow{2}{*}{ 10 } & nuc. & 26,13 & 38,21 & 31,26 & 20,16 & 12,13 & 9,11 & 5,4 & 7,5 & 32,11 & 8,15 & 5,2 \\
		& & cyt. & 6,6 & 21,11 & 6,11 & 6,7 & 7,8 & 2,10 & 8,3 & 2,2 & 14,6 & 5,4 & 1,1 \\ \vspace{-0.5\normalbaselineskip} \\
		& \multirow{2}{*}{ 20 } & nuc. & 44,18 & 32,29 & 42,29 & 12,28 & 11,11 & 12,8 & 9,7 & 8,8 & 52,24 & 14,23 & 2,3 \\
		& & cyt. & 9,10 & 25,14 & 9,12 & 6,7 & 6,8 & 15,17 & 10,5 & 4,7 & 17,10 & 3,6 & 1,1 \\ \hline \vspace{-0.7\normalbaselineskip} \\
		\multirow{6}{*}{ 4-endo } & \multirow{2}{*}{ 5 } & nuc. & 2,1 & 2,7 & 5,6 & 8,5 & 9,3 & 5,8 & 6,3 & 7,4 & 5,3 & 3,6 & 1,2 \\
		& & cyt. & 2,2 & 12,5 & 8,5 & 3,4 & 2,5 & 5,6 & 4,7 & 2,2 & 3,4 & 2,2 & 2,2 \\ \vspace{-0.5\normalbaselineskip} \\
		& \multirow{2}{*}{ 10 } & nuc. & 3,1 & 4,4 & 11,7 & 4,3 & 3,5 & 16,6 & 10,5 & 5,5 & 6,2 & 5,6 & 3,2 \\
		& & cyt. & 3,3 & 13,7 & 16,9 & 4,6 & 3,7 & 10,3 & 5,3 & 2,1 & 2,2 & 5,1 & 3,2 \\ \vspace{-0.5\normalbaselineskip} \\
		& \multirow{2}{*}{ 20 } & nuc. & 10,5 & 2,9 & 18,6 & 8,3 & 4,4 & 5,5 & 8,3 & 8,2 & 1,3 & 9,3 & 2,1 \\
		& & cyt. & 9,6 & 23,11 & 19,6 & 7,8 & 7,6 & 4,15 & 2,4 & 2,4 & 8,6 & 10,4 & 1,1 \\ \hline \vspace{-0.7\normalbaselineskip} \\
		\multirow{4}{*}{ 1-endo } & \multirow{2}{*}{ 5 } & nuc. & 13,2 & 4,3 & 2,15 & 5,4 & 7,4 & 1,5 & 2,5 & 7,6 & 6,3 & 5,5 & 5,2 \\
		& & cyt. & 1,2 & 4,5 & 9,6 & 4,6 & 7,5 & 3,5 & 3,3 & 2,4 & 2,2 & 3,2 & 3,1 \\ \vspace{-0.5\normalbaselineskip} \\
		& \multirow{2}{*}{ 10 } & nuc. & 8,1 & 5,3 & 18,7 & 3,4 & 2,5 & 15,8 & 4,4 & 10,8 & 3,14 & 1,4 & 2,2 \\
		& & cyt. & 5,2 & 10,4 & 7,9 & 0,7 & 4,5 & 4,10 & 7,2 & 1,1 & 7,2 & 2,1 & 2,1 \\ \hline \hline \vspace{-0.7\normalbaselineskip} \\
	\end{tabular}}}
\end{table}
\newpage

\subsection{DEF in a tumor with uniform macroscopic gold concentration} \label{ssec:Results-Cylinder}

\begin{figure}[htbp]
	\centering
	\includegraphics[width=0.8\textwidth]{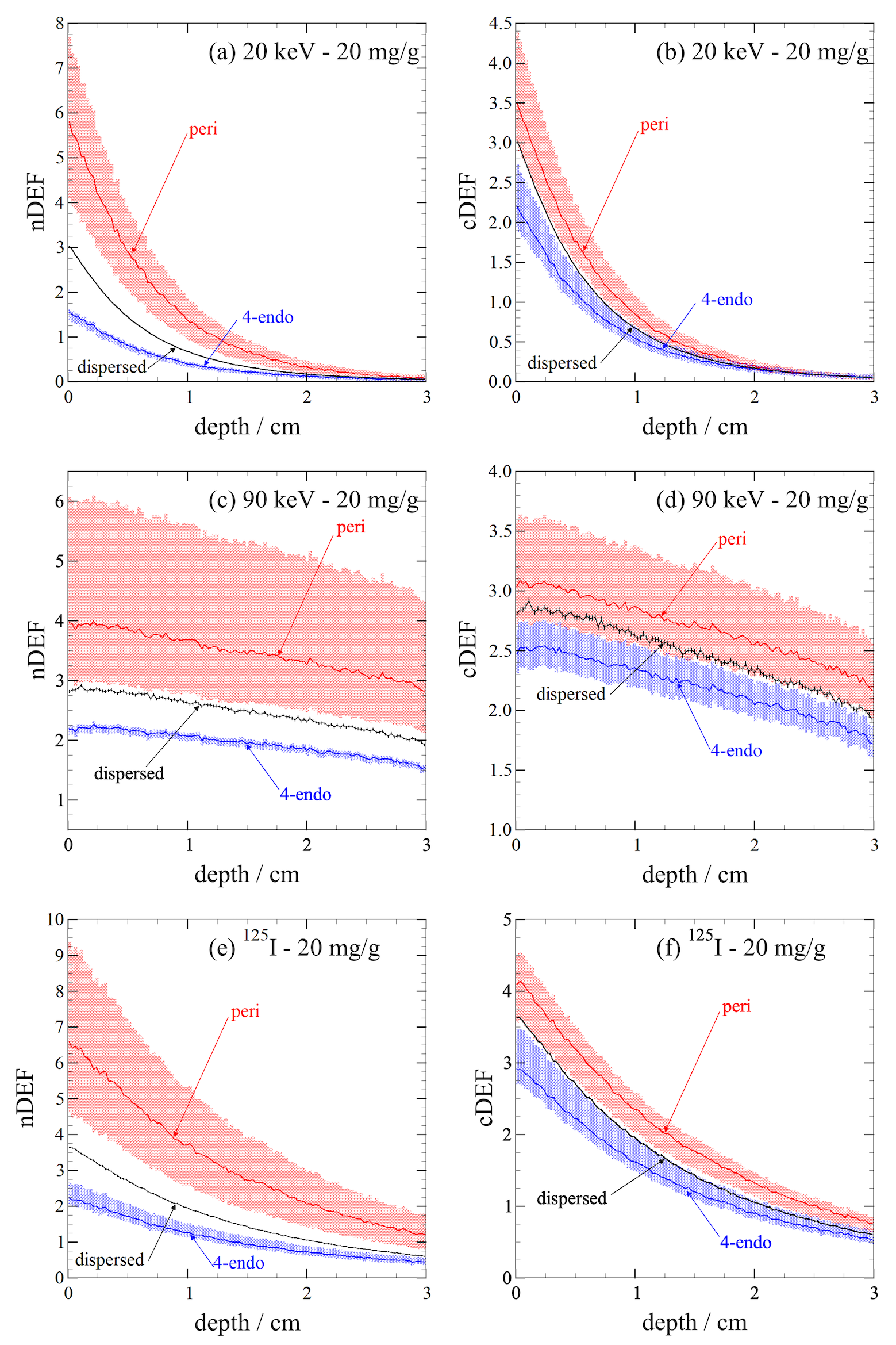}
	\captionv{\linewidth}{\cellDEFs{} as a function of depth and energy}{Nucleus (left) and cytoplasm (right) dose enhancement factor as a function of depth in a tumor of uniform macroscopic gold concentration for perinuclear (red) and four endosome (blue) configurations for 20~keV (top), 90~keV (middle), and $^{125}$I (bottom) source energies and gold concentration of 20~mg/g.  Shaded regions represent \del (Table~\ref{tab:FluxUp}) and the solid black line is tissue DEF to the scoring cavity when only modeling dispersed GNPs.
	\label{fig:cylinderResults}}
\end{figure}

Figure~\ref{fig:cylinderResults} presents \cellDEFs{}s as a function of depth at the highest gold concentrations investigated (20~mg/g) for monoenergetic 20 and 90 keV photons and an $^{125}$I source.  The colored lines represent either \peri or \fendo configurations with the corresponding shaded regions representing the intrinsic variation in DEFs given by (\delps{}, \delms{}) above. The shaded region for $^{125}$I (average energy 27.5~keV) was calculated using (\delps{},\delms{}) for monoenergetic 30~keV photons.  The black line represents dose to tissue throughout the sensitive area of the microscopic scoring cavity with 25~nm radius GNPs uniformly arranged in a hexagonal lattice geometry, the dispersed model (dDEF).  Though all DEFs are above unity for energies above 30~keV, DEFs$<$1 occur for both $^{125}$I and 20~keV energies with an nDEF below one as shallow as 3~\mm depth for the \fendo configuration at 20~keV.

The highest nDEF occurs for the \peri configuration with the $^{125}$I spectrum, ranging 4.7-9.3 at the surface and 0.9-1.7 at 3~\cm depth.  Though nDEFs with 20~keV photons are close to those with $^{125}$I near the surface (ranging 3.9-7.4), DEFs decrease more steeply as a function of depth for 20~keV energies, dropping to 0.05-0.09 after 3~\cms.  The range of nDEFs in cells with \fendo configurations vary less from the nominal values than in the \peri configuration.  At 20 keV, the \fendo configuration nDEFs are 1.52-1.68 near the surface and fall to $\sim$0.05 after 3~\cms.  The change in cDEF as a function of energy is similar to the nDEF trends, though \peri cDEFs are the same or lower than their equivalent nDEFs and \fendo cDEFs are the same or higher than nDEFs.  For all cases, the largest expected \fendo \cellDEFs{}s are below the lowest expected \peri \cellDEFs{}s.

For the photon energies for which results are not shown in Figure~\ref{fig:cylinderResults}, the change in DEFs with energy on a macroscopic scale follow previous results\cite{MT17} simulated with no cell model and on a microscopic scale exhibit similar trends to those found in Part~I.  The differences between them are not drastic for energies  between 40 and 80~keV or over 90~keV. \cellDEFs{}s for $^{131}$Cs and $^{103}$Pd, as well as 30~keV, were very close to the $^{125}$I \cellDEFs{}s due to the similar photon energies.

\begin{figure}[htbp]
	\centering
	\includegraphics[width=0.8\textwidth]{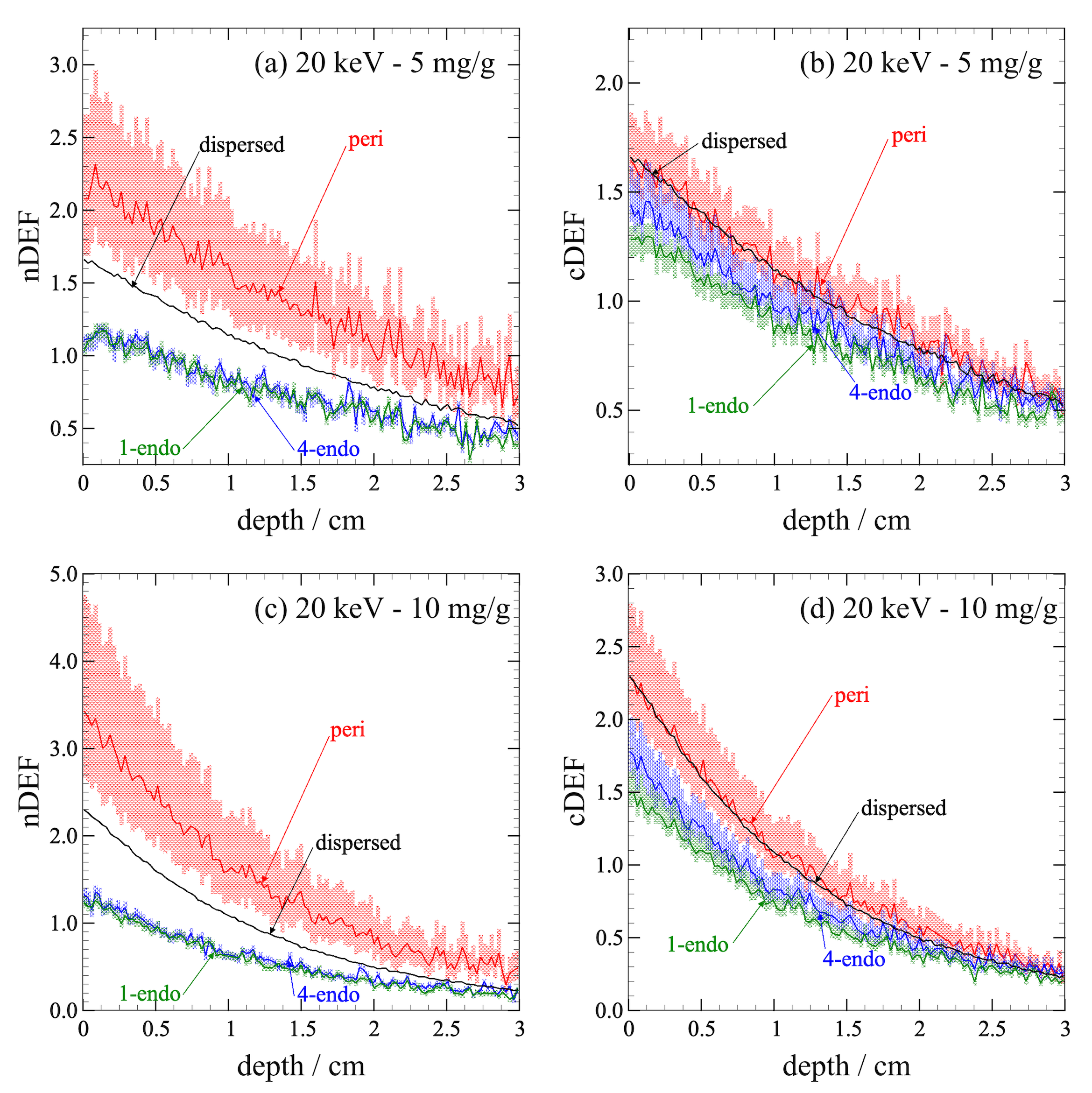}
	\captionv{\linewidth}{\cellDEFs{} as a function of depth and gold concentration}{Nucleus (left) and cytoplasm (right) dose enhancement factors as a function of depth in a tumor of uniform macroscopic gold concentration for perinuclear (red), four endosome (blue) and single endosome (green) configurations for 20~keV photons and gold concentrations of 5 (top) and 10~mg/g (bottom).  Shaded regions represent \del (Table~\ref{tab:FluxUp}).
	\label{fig:cylinderResults2}}
\end{figure}

Figure~\ref{fig:cylinderResults2} shows \cellDEFs{}s for a 20~keV source energy at concentrations of 5 and 10~mg/g.  The \peri configuration has the highest \cellDEFs{},  as in the 20~mg/g case.  Though the DEFs near the surface at 5 and 10~mg/g are lower than the analogous surface DEFs at 20~mg/g, they become relatively larger than their analogs at depths between 0.2-1.0~cm due to the lower gold content in the scatter media.  The \oendo and \fendo configuration nDEFs agree within uncertainty at all depths; when comparing cDEFs of the two configurations, the \oendo configuration cDEFs are lower than the \fendo cDEFs at the same depth largely due to the increased absorption within the larger endosome - recall that the same mass of gold is contained in a single endosome for the \oendo case and distributed between four endosomes in the \fendo case. 

For all scenarios, dDEFs are below \peri and above \fendos{} \cellDEFs{}s. \fendos{}  \cellDEFs{}s are generally the same or steeper than dDEFs, and \peri \cellDEFs{}s are generally the same or less steep than dDEFs. However, if the ratios of \cellDEFs{}s to dDEFs are taken (not shown), the ratios are generally approximately constant with depth; thus, the relative differences in DEF between models does not vary with depth.

\subsection{DEF in a tumor with spatially-varying gold concentration} \label{ssec:Results-Sphere}

\begin{figure}[htbp]
	\centering
	\includegraphics[width=0.4\textwidth]{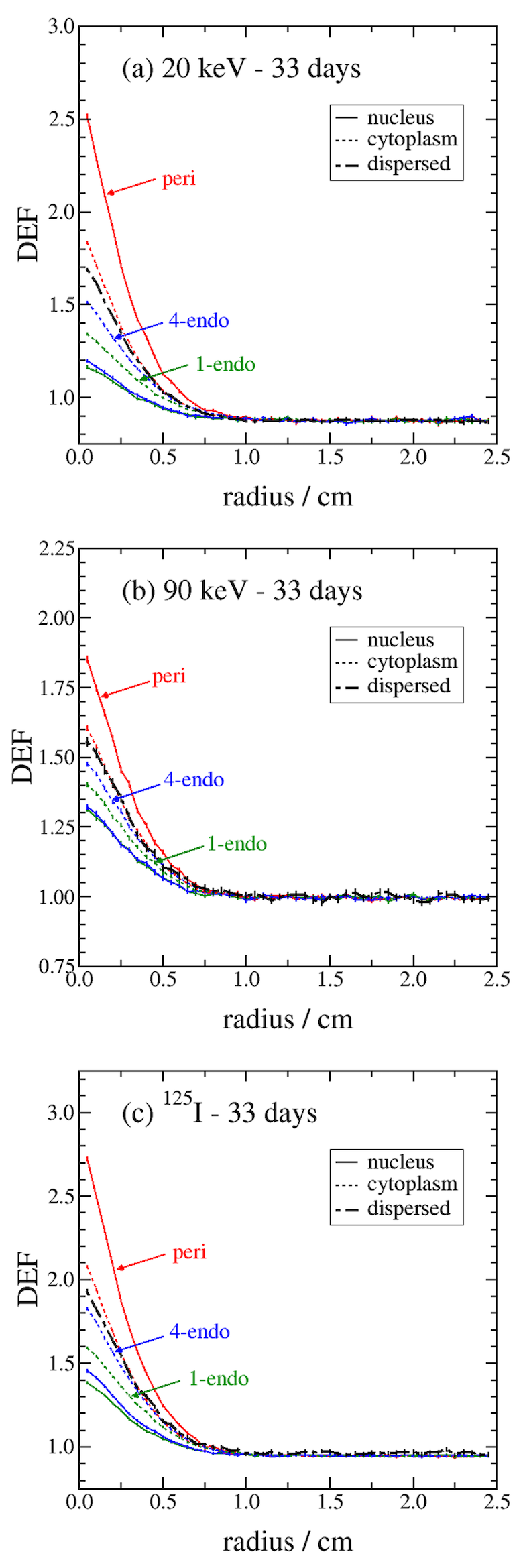}
	\captionv{0.95\linewidth}{\cellDEFs{} in volume after 33~days}{\cellDEFs{} as a function of radius in a sphere of spatially-varying gold concentration after 33~days of elution containing a (a) 20~keV, (b) 90~keV, and (c) $^{125}$I point source in its centre for \peris, \oendos, and \fendo configurations.  Black line is tissue DEF to the scoring cavity when modeling only dispersed GNPs.
	\label{fig:sphereResults}}
\end{figure}

Figure~\ref{fig:sphereResults} shows \cellDEFs{}s of the three cell GNP configurations (\peris, \oendos, and \fendos) and dDEFs as a function of radius for 20~keV, 90~keV, and $^{125}$I energies in a sphere containing gold eluted from the center over 33~days, with dose calculated on day 33.  Results for 5 and 59 days follow similar trends and were thus omitted.  Results for 1 and 200 days either had too little dose enhancement or too much attenuation due the gold, respectively, and were omitted as well.  The 33~day results for the dispersed 25~nm radius GNPs model with $^{125}$I in this paper are close to the 6711 seed results computed previously (Figure~6a in the previous work\cite{MT17}).

All of the GNP configuration models (and the dispersed model) converge to DEFs of $\sim$0.86, $\sim$0.99, and $\sim$0.94 at radii $>$1~\cm for 20~keV, 90~keV, and $^{125}$I, respectively.  This convergence occurs in the regions where the gold concentration is essentially zero and the choice of model in the microscopic scoring cavity no longer matters, as it is all modeled as tissue; the convergence occurs at lower radii for 1 and 5~days and at larger radii for 59 and 200~days, as expected based on gold concentration (results not shown).  Thus, the choice of model in the microscopic scoring cavity does not affect primary photon attenuation, verifying that the models are equivalent on the macroscopic scale.  For the point nearest the centre (r=0.05~\cms), the highest nDEF (2.9) and cDEF (2.2) are found for the \peri configuration for the $^{125}$I spectrum and the lowest central nDEF (1.3) and cDEF (1.4) are for the \oendo and \fendos, respectively, for 90~keV photons. 

\section{Discussion} \label{sec_GNPT:Discussion}

Part~I of this study established \cellDEFs{} variations with cell size, intracellular GNP configuration, gold concentration, and incident photon energy. This work, Part~II, builds on those results to assess intrinsic variations in \cellDEFs{}s with local fluctuations in gold concentration and cell size, and uses the HetMS framework to model \cellDEFs{}s within tumors in order to assess the factors at play across these macroscopic volumes.

A thirteen cell model is used to compute the variation of \cellDEFs{}s with biological fluctuations in cell/nucleus radius and target/neighboring cell gold concentration, represented by \delp and \delms, the upper and lower DEF variations, respectively.  The \delm values are of particular interest as they represent the lower end of DEF values one might expect across a population of cells (lower than the mean by as much as 29\% in some cases), which are relevant for both developmental research and clinical outcomes of GNPT. The combined \del values (Table~\ref{tab:FluxUp}) provide (at least) a first order approximation of the intrinsic variation of GNPT dose enhancement on a cellular level\cite{Ch10b,CC07,Vo00}.  Although a 20\% fluctuation in gold concentration could be a large underestimate of the actual difference in cell-to-cell GNP uptake\cite{Ja04a}, the 20\% variation was chosen to provide an (ideal) representative value for proposed GNPT deliveries/treatments in the literature\cite{As15,Br17,Si15,Ci15}. These \del values can be used to extrapolate \cellDEFs{}s for a reference cell to expected ranges of \cellDEFs{}s over an entire population of tumor cells in a GNPT scenario. Figure~\ref{fig:cylinderResults} exemplifies this, showing the ranges of \cellDEFs{}s expected in a simple GNPT scenario as a function of depth for different photon energies and demonstrating the large differences (up to 100\%) in \cellDEFs{}s that can occur over a cell population at the same distance from the source.

The HetMS simulations of GNPT presented herein allow for efficient single simulation calculation of \cellDEFs{}s in complex geometries, a cylinder with a spatially-uniform distribution of GNPs and a sphere with a spatially-varying distribution of GNPs with radius. The \cellDEFs{}s at the surface of the spatially-uniform model (Figure~\ref{fig:cylinderResults}) at 20~keV are within a few percent of the equivalent reference \cellDEFs{}s in Figure~4 of Part~I.  Surface \cellDEFs{}s at 90~keV, on the other hand, are markedly larger in the spatially-uniform simulation; this increase is due to the increased number of scattered photons in the larger volume entering the microscopic scoring cavity.  Figure~\ref{fig:cylinderResults} demonstrates that \cellDEFs{}s decrease with depth and can fall below one. Sub-unity DEFs for $^{125}$I were also found in the homogeneous mixture DEF values calculated in the prostate simulations of Brivio \etal\cite{Br17}, with DEFs as low as 0.7 in parts of the treatment volume.  In the model with GNPs dispersed evenly in tissue, dDEFs lie between the \peri and \fendo \cellDEFs{}s in all cases.  The dDEFs are larger than the endosome \cellDEFs{}s due to the GNPs being spread evenly throughout the scoring volume, minimizing photoelectron absorption in neighbor GNPs.  dDEFs are smaller than \peri \cellDEFs{}s because, despite the photoelectron absorption in gold in the \peri configuration, the GNPs are positioned very close to both scoring volumes due to the thinness of the gold-containing region.

As in Part I, \cellDEFs{} results for all HetMS simulations in this work show that the \peri configuration is the ideal configuration for maximizing dose.  This is primarily because GNPs in the \peri configuration are not clustered, thus generated photoelectrons leave the gold-containing region having lost little energy to the GNPs\cite{Ki17}.  Therefore biological delivery methods that avoid GNPs clustering into a compartment within the cell are ideal for higher nucleus \textit{and} cytoplasm DEFs.  In contrast, biological delivery methods that result in GNPs clustering into one or more endosomes can result in nDEF up to only 1.5 in the 0.25~\cm closest to the source (Figure~\ref{fig:cylinderResults}a) with DEF dropping below unity at farther depths; this delivery method (if it achieves the same GNP uptake in cells) is therefore undesirable for any GNPT aiming to increase nucleus dose within even 1~\cm of the treatment source(s).

For the tumor with a spatially-uniform gold distribution, \cellDEFs{}s and dDEFs generally follow the same trends with depth, with the ratios of \cellDEFs{}s to dDEF approximately the same at $0\,$cm and at $3\,$cm for each GNP configuration. This means that for a tumor with uniform gold concentration, a simple multiplicative factor is needed to convert dDEF to \cellDEFs{}s for different GNP configurations. In the tumor with spatially-varying gold distribution, however, this is no longer the case. DEFs for all of the configurations (including dispersed) differ close to the source but converge to the same value after approximately $1\,$cm, meaning the ratios of of \cellDEFs{}s to dDEF changes with depth. This change is due to different positions of GNPs relative to scoring volumes between the cell and dispersed models. In the cell models, the gold-containing regions are at fixed positions in the cells at a set distance from the scoring volumes (nucleus and cytoplasm), thus, the distance from any part of the scoring volume to the nearest GNP does not change with concentration, but rather the number of GNPs in the cell does.  Conversely, in the dispersed model, at low concentrations GNPs are spaced much farther apart than in the cell case and the distance between two neighboring GNPs further increases as concentration decreases.  For very low concentrations, a large fraction of a scoring volumes nearest GNPs are beyond the expected photoelectron range\cite{Jo10}, causing a further decrease in \cellDEFs{}s due to the lower number of photoelectrons generated at low concentrations (the primary contributor to \cellDEFs{} decrease in the cell models). 

In both Parts~I and~II of this work, we have generally limited our investigations to cells containing gold either in a layer surrounding the nucleus or in one or four endosomes. In reality, GNP uptake is not this precise and might occur as a combination of these configurations\cite{Ch10d,Li08d}. In Part~I we established that for the single cell scenario, dose enhancement for a cell with a combination of configurations can be predicted using the fits to the individual configurations. Future work can extend these considerations to the far more complex tumor-scale scenarios presented here in Part~II. In addition, we have presented our results as a function of nucleus and cytoplasm dose, which does not fully capture all aspects of GNP radiosensitization\cite{Bu12b,Ro17}. Future work would include translating dosimetric results to biological effects, taking into account realistic GNP configurations. Such work might make use of the  local effect model (LEM) which attempts to relate energy deposition within a cell to the survival probability of that cell\cite{Mc11a,Mc11b,BC17,Le13}. The use of these models would advance comparisons of computational results to experimental ones and may help to bridge any differences seen between the two types of results.

\section{Conclusion} \label{sec_GNPT:Conclusion}

This paper presents HetMS simulations of GNPT that account for both the macroscopic effects of introducing gold to a treatment volume, as well as the microscopic effects of GNPs on a cellular level.  In addition, intrinsic variation of \cellDEFs{}s (due to fluctuations in local gold concentration and variations in cell/nucleus size) represent estimates of the range of DEFs over a population of cells (rather than a single DEF for a nominal cell), with the lower bound representing \cellDEFs{}s relevant for tumor control considerations.  Similar to the single cell results of Part~I, the large surface area and proximity to scoring volumes of GNPs in the perinuclear configuration lead to the largest \cellDEFs{}s when compared to the endosome gold configurations.  

In some optimal cases with little variation in the gold distribution, conversion factors might be applied to simplified simulations, such as the dispersed scenario presented here, in order to calculate \cellDEFs{}s across the treatment volume. In more complex (and realistic) cases, however, this simplified method breaks down and detailed simulations of individual cells and GNPs over the whole treatment volume (and potentially nearby healthy tissue) are required for accurate dosimetric evaluations. Performing these full patient, multiscale simulations can take prohibitively long times when using more na{\"i}ve simulation approaches.  Fortunately, as evidenced in this study,  the Heterogeneous MultiScale approach used herein is well suited to this task, and is a promising framework for future GNPT planning and research.

Collectively, Parts~I and II have shown that to plan for GNPT treatment, there are a multitude of factors at play that need to be accounted for, including cell and nucleus radius, GNP distribution over the entire treatment volume, beam quality, and GNP configurations within the cell, which have all been explored in these works.  In addition to these factors, the interplay between them can vary with position in the tumor; when computing nominal cell DEFs at different positions, the variation over a population of cells should be accounted for.  Care must be taken to choose an appropriate computational model for the GNPT scenario being considered in order to achieve accurate results.

\section*{Acknowledgments}

The authors acknowledge support from the Kiwanis Club of Ottawa Medical Foundation and Dr. Kanta Marwah Scholarship in Medical Physics, the Natural Sciences and Engineering Research Council of Canada (NSERC) [funding reference number 06267-2016], Canada Research Chairs (CRC) program, an Early Researcher Award from the Ministry of Research and Innovation of Ontario, the Queen Elizabeth II Graduate Scholarships in Science and Technology, the Ontario Graduate Scholarship, and the Carleton University Research Office, as well as computing support provided by the Shared Hierarchical Academic Research Computing Network (SHARCNET, \href{https://www.sharcnet.ca}{sharcnet.ca}) and the Digital Research Alliance of Canada (\href{https://www.alliancecan.ca}{alliancecan.ca}). 

\section*{Conflict of interest}

The authors have no conflicts to disclose.


\setlength{\baselineskip}{0.55cm}

\bibliography{GNPT_Paper_final.bbl}

\bibliographystyle{medphy.bst}

\end{document}